\documentclass[cameraready]{Interspeech}

\title{Probabilistic Verification of Voice Anti-Spoofing Models}

\author[affiliation={1,2,3}, equalcontribution]{Evgeny}{Kushnir}
\author[affiliation={4},  equalcontribution]{Alexandr}{Kozodaev}
\author[affiliation={1,5}, equalcontribution,correspondingauthor]{Dmitrii}{Korzh}
\author[affiliation={1,6}, equalcontribution]{Mikhail}{Pautov}
\author[affiliation={7}]{Oleg}{Kiriukhin}
\author[affiliation={1,3,5}]{Oleg Y.}{Rogov}

\address{
    $^1$AXXX,
    $^2$HSE,
    $^3$Applied AI Institute,
    $^4$Central University,
    $^5$MTUCI,
    $^6$Trusted AI Research Center, RAS,
    $^7$City University of Hong Kong
}

\email{d.s.korzh@mtuci.ru}

\keywords{voice anti-spoofing, audio deepfake detection, robustness, verification, text-to-speech, voice cloning}

\usepackage{comment}

\usepackage{graphicx} 
\usepackage{xcolor}
\usepackage{amsfonts}
\usepackage{dsfont}
\usepackage{booktabs}
\usepackage{makecell}
\usepackage{multirow}
\usepackage{algorithm}
\usepackage{algpseudocode} %
\usepackage{amsmath,amssymb}
\usepackage{soul}

\begin{document}

\maketitle

\begin{abstract}
    Recent advances in generative models have amplified the risk of malicious misuse of speech synthesis technologies, enabling adversaries to impersonate target speakers and access sensitive resources. Although speech deepfake detection has progressed rapidly, most existing countermeasures lack formal robustness guarantees or fail to generalize to unseen generation techniques. We propose PV-VASM, a probabilistic framework for verifying the robustness of voice anti-spoofing models (VASMs). PV-VASM estimates the probability of misclassification under text-to-speech (TTS), voice cloning (VC), and parametric signal transformations. The approach is model-agnostic and enables robustness verification against unseen speech synthesis techniques and input perturbations. We derive a theoretical upper bound on the error probability and validate the method across diverse experimental settings, demonstrating its effectiveness as a practical robustness verification tool.
\end{abstract}

\section{Introduction}
Over the past decade, rapid advancements of text-to-speech (TTS) and voice cloning (VC) models~\cite{casanova2024xtts,azzuni2025voice}  resulted in both improved generated speech quality and notably easier access to it for a broad audience. A variety of online and offline solutions, while generally beneficial, pose serious security risks~\cite{wang2025one}: in particular, realistic synthetic speech can be exploited to impersonate target speakers and gain unauthorized access to sensitive resources. To counter this threat, the research on voice anti-spoofing (VAS) and speech deepfake detection substantially accelerated in recent years. 

Primary spoof detection methods are based on feature engineering, architectural design, training optimization techniques, and demonstrate quantitative efficiency in several data-dependent scenarios \cite{li2025survey}. 
Despite notable empirical progress, modern voice anti-spoofing models remain non-robust in practice: they are usually deployed in conditions that notably differ from those they are expected to be robust to. Results of relevant competitions, such as ASVspoof~\cite{wang2024asvspoof} and Int-the-Wild~\cite{muller2022does}, confirm this statement,  demonstrating that state-of-the-art models suffer from a significant performance degradation when exposed to previously unseen spoof generation methods or new audio conditions and domains. Consequently, strong empirical accuracy alone provides a limited sense of reliability in real-world applications.

A fundamental challenge underlying this limitation is the absence of formal robustness guarantees~\cite{li2023sok}. Existing VAS measures are predominantly evaluated empirically and offer no principled bounds on their behavior under perturbations or generative processes. While the broader machine learning literature has developed a rich body of robustness certification methods, these techniques are typically tailored to a narrow class of perturbations and are not directly applicable to the complex, generative transformations induced by modern speech generators. As a result, the certification of VAS models against TTS, VC, or other neural speech synthesis systems remains largely unexplored.

In this paper, we bridge this gap by proposing \emph{PV-VASM}, a probabilistic framework for robustness verification of voice anti-spoofing models in a black-box model-agnostic manner. PV-VASM yields an upper bound on the probability of incorrect classification of the conventionally transformed or artificially synthesized input audio. The proposed framework enables verification against unseen transformation and speech generators, making it relevant for real-world pre-deployment robustness evaluation. 

From a technical point of view, our approach builds on probabilistic concentration inequalities and provides a principled mechanism for estimating tight upper bounds on misclassification probabilities with high confidence. We develop practical procedures for estimating the statistics of random variables required to estimate the bound and for selecting certification parameters to balance the tightness of the results versus computation cost. Extensive experiments across a wide range of transformations, TTS, and VC models demonstrate that PV-VASM yields meaningful robustness certificates and complements standard empirical evaluation.

\textbf{The contributions of this work might be summarized as follows:}
\begin{itemize}

    \item We introduce and motivate the probabilistic framework to formally verify the robustness of voice anti-spoofing models.
    We propose PV-VASM, a model-agnostic method capable of verifying robustness not only against classic audio transformations but also against arbitrary neural speech generators, including unseen TTS and VC systems. 
    \item We derive a theoretical upper bound on the error probability of the method and present a practical pipeline for estimation of statistics of underlying random variables and certification parameters.
    \item We empirically validate the proposed framework on diverse experimental settings, demonstrating its practical applicability and relevance to real-world deployments of voice anti-spoofing models.
\end{itemize}

Overall, PV-VASM provides a systematic approach to verify the robustness of voice anti-spoofing models, particularly in the context of rapidly advancing speech synthesis technologies.

\section{Related work}

Neural networks are well known to exhibit performance instability under domain shift~\cite{10017290}, often resulting in limited generalization beyond the training distribution. Similar behavior has been observed for voice anti-spoofing (VAS) models in major evaluation campaigns, including the ASVspoof challenges~\cite{todisco2019asvspoof, wang2020asvspoof, yamagishi2021asvspoof, wang2024asvspoof} and the ADD series~\cite{yi2022add, yi2023add}. In these benchmarks, systems typically demonstrate a notable degradation in test performance in comparison to validation performance, as the evaluation data introduces the new perturbation types and speech generation methods that are not presented in the training set.

Common VAS architectures include graph-based neural approaches, such as AASIST~\cite{jung2022aasist}, which construct both homogeneous and heterogeneous graphs and apply graph attention mechanisms to structures derived from temporal and spectral representations (e.g., Sinc convolutions~\cite{ravanelli2018speaker}). Substantial performance gains have been achieved by leveraging self-supervised learning (SSL) audio encoders, including Wav2Vec~2.0~\cite{baevski2020wav2vec} and WavLM~\cite{chen2022wavlm}, as front-end feature extractors. These features can subsequently be processed by a variety of back-end architectures, such as AASIST (e.g., Wav2Vec2--AASIST~\cite{tak2022automatic}) or simpler pooling-based models followed by linear classifiers~\cite{aliyev2024intema}. Additional performance improvements can be achieved through dataset expansion and diversification. Further gains have been reported via model ensembling, one-class learning paradigms, the use of specialized architectural components and loss functions, as well as the integration of audio large language models~\cite{foret2020sharpness, ding2023samo, borodinaasist3, gu2025allm4add}. Despite these advances, VAS systems frequently exhibit limited robustness to previously unseen spoofing generators, resulting in substantial performance degradation in open-set evaluation scenarios. Moreover, such models remain vulnerable to adversarial manipulations and synthetic input perturbations, including additive noise and signal-level transformations~\cite{rabhi2024audio, li2025measuring}.

In recent years, the vulnerabilities of deep learning models have been extensively studied, and the vulnerability of models to different types of input perturbations has been demonstrated~\cite{sehwag2019analyzing, croce2020robustbench, hendrycks2019benchmarking}. In particular, adversarial attacks~\cite{szegedy2013intriguing, goodfellow2014explaining} have received substantial attention due to their ability to induce severe performance degradation with imperceptible input modifications. In response, a variety of defense mechanisms have been proposed to enhance model robustness, which can be broadly classified into empirical and certified approaches. Empirical defenses~\cite{wong2020fast, madry2017towards, fort2024ensemble}, such as adversarial training, are comparatively easy to deploy and have demonstrated strong empirical effectiveness. Nevertheless, these methods do not provide formal guarantees against previously unseen or adaptive perturbations, often giving rise to a recurring arms race between attack and defense strategies. 

To address this limitation, the certification paradigm has been introduced~\cite{lecuyer2019certified, li2023sok}, offering provable deterministic or probabilistic guarantees on a model’s behavior under all perturbations within a predefined threat model. These perturbations are typically characterized by the type of transformation and their parameters (for example, norm-bounded additive perturbations \cite{goodfellow2014explaining}).  A comprehensive survey of certified robustness methods is provided in~\cite{li2023sok}. Despite this progress, limited attention is devoted to the certification of voice anti-spoofing (VAS) systems in the literature. In particular, existing certification techniques are generally incapable of verifying robustness to synthetic speech generation, a crucial property for the safe deployment of VAS models in real-world scenarios.

Among existing certification techniques, probabilistic methods~\cite{mohapatra2020higher, tit2021efficient, baluta2021scalable, pautov2022cc} estimate the probability that a given input is adversarial and appear particularly suitable for application in the voice anti-spoofing setting. For instance, CC-Cert~\cite{pautov2022cc} provides an upper bound on the misclassification probability of a transformed input, where the transformation parameters are sampled from a prescribed distribution. This bound is derived using an empirical form of the Chernoff-Cramer concentration inequality~\cite{boucheron2003concentration}. 
However, such approaches do not offer robustness guarantees for non-analytical transformations, including data generated by artificial or neural speech generators. Furthermore, in these methods, theoretical formulation implies the knowledge of the exact values of statistics of the distribution of the input perturbations, such as the coefficient of variation. An assumption about the invariance of these values across different possible transformations is generally violated and may lead to overly conservative robustness estimates.

\section{Methodology} 

In this section, we provide the formal description of the proposed method. We recall that PV-VASM is designed to certify the probability of the voice anti-spoofing model to misclassify the transformed input audio. The following section is devoted to the verification of robustness when the input audio is subjected to a parametric label-preserving transformation; in the subsequent sections, we discuss the verification of the model's robustness against the speech generation methods. Here and below, we treat the VAS problem as a binary classification task.

\subsection{Problem setup}

Let $\mathcal{X} \subset \mathbb{R}^d$ be the space of input audios, $x\in \mathcal{X}$ be the fixed input audio, ${f}:\mathbb{R}^d \to \mathbb{R}^2$ be the source voice anti-spoofing model that assigns the vector of probabilities to $x$ in the form
\begin{equation}
    f(x) = (p_1, p_2)^\top, 
    \end{equation}
where $p_1+p_2=1,\ p_1,p_2\ge0.$ Here, $p_1$ represents the probability of $x$ being classified as a spoof audio, and $p_2$ represents the probability of $x$ being classified as a bona fide audio. To ease the notation, we introduce the classification rule in the form 
\begin{equation}
    h(x) = \arg\max_i f_i(x).
\end{equation}
In this work, we focus on the problem of certifying the robustness of ${f}$ under parametric perturbations of its input. Formally, the parametric transformation $\phi: \mathcal{X} \times \Theta \to \mathcal{X}$ is the mapping of input audio space to itself, where $\Theta$ is a fixed space of parameters of the transformation $\phi.$  Under transformation $\phi,$ the input object $x$ becomes the random variable, since it is  a function of $\theta$:
\begin{equation}
    x' = \phi(x, \theta), \quad  \theta \overset{\mu}{\sim} \Theta, 
\end{equation}
where $\mu$ is the measure on $\Theta$. 
The prediction of the source model on the perturbed audio becomes the random variable in the form
\begin{equation}
\label{eq:aug_pred}
    {f}(x') = (p_1', p_2')^\top.
\end{equation}
The robustness of ${f}$ under label-preserving transformation $\phi$ at point $x$ is reflected by the probability to assign the same class index to $x$ and $x'$ in the form
\begin{equation}
\label{eq:main_prob}
    \mathbb{P}_{\theta \sim\Theta} \left[h(x) = h(x') \right], \ \text{where }x' = \phi(x,\theta).
\end{equation}
Here we omit $\mu$, assuming it is fixed and known. 
In the case of a binary classification task, 
\begin{align}
\label{eq:clf}
    h(x) = h(x') & \leftrightarrow\ \left(p_1 - 1/2\right) \left(p_1' - 1/2\right) > 0  \nonumber \\ & \leftrightarrow\left(p_2 - 1/2\right) \left(p_2' - 1/2\right) > 0.
\end{align}
Without loss of generality, we assume that the initial audio $x$ is correctly classified by $f$ as a bona fide one, so $p_2 > 1/2$. Thus,
\begin{equation}
    h(x) = h(x') \leftrightarrow p_2' > 1/2
\end{equation}
and 
\begin{equation}
\label{eq:second_main_prob}
    \mathbb{P}_{\theta \sim\Theta} \left[h(x) \ne h(x') \right] = \mathbb{P}_{\theta \sim\Theta} [p_2' < 1/2].
\end{equation}
In case of a nontrivial $\mu$ and $f$, the probability from Eq.~\eqref{eq:second_main_prob} is intractable. The goal of the proposed method is to provide a tight upper bound for this probability and, consequently, a tight upper bound for the probability of $f$ to misclassify $x'.$

\subsection{Description of PV-VASM}

To ease the notation, we introduce the random variable $Z \equiv p_2'.$ Then, according to Chernoff inequality, 

\begin{equation}
    \label{eq:prob_bound}
    \mathbb{P}_{\theta \sim \Theta} [Z < 1/2] \le \inf_{t<0} \mathbb{E}(e^{tZ}) e^{-t/2}. 
\end{equation}
Since the expectation from Eq.~\eqref{eq:prob_bound} is intractable, in our method, it is upper-bounded via sampling the random variable $Z$ and computing the statistics of the resulting sample \cite{pautov2022cc}. Specifically, for the given $t<0,$ let 
\begin{align}
\label{eq:sample_mean}
    Y_j = \frac{1}{n} \sum_{i=1}^n \exp(tZ^j_i) \exp(-t/2), \ j \in [1,\dots,k]
\end{align}
be the set of $k$ independent and identically distributed sample means, each computed over $n$ realizations of $Z^j$ (here, superscript $j$ denotes the batch of realizations of the random variable used to compute $Y_j$). Then, for all $\delta \in (0,1)$, the statistic 
\begin{equation}
\label{eq:output}
    \mathcal{A}(x) =\max \left\{Y_1,\dots,Y_k\right\} \delta^{-1} 
\end{equation}
is an upper bound for the expectation $\mathbb{E}(e^{tZ})e^{-t/2}$ with high probability. 
Specifically, the error probability of the method is upper-bounded by 
\begin{align}
 \label{eq:cc_cert_error}
    \mathbb{P}_{\theta \sim \Theta} \left[\mathcal{A}(x) <  \mathbb{E}(e^{tZ})e^{-t/2}\right] \le & \left(\frac{1}{1 + n (1-\delta)^2c^{-2}}\right)^k \equiv \nonumber \\
    \equiv p(n,k,c),
\end{align}
where 
\begin{equation}
\label{eq:c}
    c = \frac{\sqrt{\mathbb{V}(e^{tZ})}}{\mathbb{E}(e^{tZ})}
\end{equation}
is the coefficient of variation of the random variable $e^{tZ}$ (see~\cite{pautov2022cc}). Informally, given $m = n\times k$ realizations of a random variable $Z$, the statistic $\mathcal{A}(x)$ provides an upper-bound for the probability of misclassifying $x'$, that holds with high probability. 

In the next section, we describe the procedure to estimate the coefficient of variation from Eq.~\eqref{eq:c} to compute the error probability of the method. 

\subsubsection{Estimation of error probability} 

PV-VASM yields an upper bound for the probability from Eq.~\eqref{eq:second_main_prob} in the form of Eq.~\eqref{eq:output}. Here and below, we say that PV-VASM \emph{makes an error} if its output underestimates an unknown error probability, namely, if
\begin{equation}
\label{eq:error}
    \mathbb{P}_{\theta \sim \Theta } [Z <1/2] > \mathcal{A}(x). 
\end{equation}
To estimate the probability of error from Eq.~\eqref{eq:error}, one has first to estimate the right-hand side of Eq.~\eqref{eq:cc_cert_error}. Since both the mean and variance of the random variable $e^{tZ}$ are intractable, one can estimate the coefficient of variation using $m=n\times k$ realizations of the random variable $Z$. In our approach, we use one-sided confidence interval estimation of the coefficient of variation in the form of the modified McKay's approximation \cite{payton1996confidence}:
\begin{equation}
\label{eq:c_error}
    \mathbb{P}_{\theta \sim \Theta} \left[c > \left(\frac{\chi^2_{\alpha/4} (1+\hat{c}^2)}{m\hat{c}^2}\right)^{-1/2}\right] < \alpha/2.
\end{equation}
Here $\alpha \in (0,1)$ is the desired confidence level, $\chi^2_{\alpha/4}$ is the lower  $\alpha /4-$percentile of the Chi-square distribution with $m-1$ degrees of freedom, and $\hat{c}$ is the sample coefficient of variation. 


Note that by combining Eq.~\eqref{eq:cc_cert_error} and Eq.~\eqref{eq:c_error}, one can estimate the probability of an error of PV-VASM. Specifically, by introducing an auxiliary random variable 
\begin{align}
\label{eq:indicator}
        A = \mathds{1} \left\{\mathbb{P}_{\theta \sim \Theta } [Z <1/2] > \mathcal{A}(x)\right\},
\end{align}
and setting $\tilde{c} = \left({\chi^2_{\alpha/4} (1+\hat{c}^2)m^{-1}\hat{c}^{-2}}\right)^{-1/2}$ from Eq.~\eqref{eq:c_error},
one can upper bound the probability from Eq.~\eqref{eq:second_main_prob} by  
\begin{align}
\label{eq:gurantees}
    \mathbb{P} [A=1] & < 1 \times  \mathbb{P}[c > \tilde{c}] +  \nonumber 
    \\ & + p(n,k,c=\tilde{c}) \mathbb{P}[c \le \tilde{c}] & < \nonumber \\
    &< \alpha/2 + p(n,k,c=\tilde{c}).
\end{align}
When the number of samples is large enough, so that  $p(n,k,c=\tilde{c}) < \alpha/2$, the misclassification probability from Eq. ~\eqref{eq:second_main_prob} is upper bounded by $\alpha$, since $\mathbb{P} [A=1]  < \alpha$. 

It is worth noting that for sufficiently large values of $c$, McKay’s approximation in the form from Eq.~\eqref{eq:c_error} should be replaced by, for example, bootstrap-based interval estimation. 
Recall that the definition of $\mathbb{P} [A=1] $ from Eq.~\eqref{eq:indicator} is given for the verification of an initially bona fide audio. To verify robustness to the transformations of initially spoof audio, we flip the sign of parameter $t$ from Eq.~\eqref{eq:prob_bound}: $t \mapsto -t$.

\begin{algorithm}[tbh]
\caption{PV-VASM, the case of input transformations}
\label{alg:certification}
\begin{algorithmic}[1]
\Require Classifier $f$, verification dataset $\mathcal{D}$, transform $\phi$ and parameter space $\Theta$,  hyperparameters $n,k,\delta,\alpha,\varepsilon$
\Ensure Probabilistically Certified Accuracy $\mathrm{PCA}(\varepsilon, \alpha, \mathcal{D})$
\State $S \gets 0$ 
\For{$(x,y)\in \mathcal{D}$}
    \State $Z \in \mathbb{R}^{k\times n} \gets \textsc{AugmentPredict}(f,\phi,\Theta,x,n,k)$
    \State $q \gets \mathds{1} \left\{h(x)=y \right\}$

    \State $\mathcal{A}(x) \gets  \min_{t} \max_{j\in[k]} Y_j(Z[k])$
    \Comment{according to Eq. \eqref{eq:sample_mean}}
    \State $t^{*} \gets \arg [\min_{t} \max_{j\in[k]} Y_j(Z[k])]$
    \Comment{save the best value of $t$}
    \State $\mathcal{A} (x) \gets  \mathcal{A} (x)  / \delta$
    \State $\tilde c \gets \text{EstimateC}(Z, t^{*},\alpha/2)$ \Comment{one-sided interval estimation according to Eq. ~\eqref{eq:c_error}}
    \State $p \gets \left(1 + n(1-\delta)^2/\tilde c^2\right)^{-k}$
    \State $S \gets S + \mathds{1}\left\{q \wedge \mathcal{A}(x)< \varepsilon\wedge  \ p < \alpha/2 \right\}$
\EndFor
\State $\mathrm{PCA}(\varepsilon,\alpha, \mathcal{D}) \gets S / |\mathcal{D}|$
\State \Return $\mathrm{PCA}(\varepsilon,\alpha, \mathcal{D})$
\end{algorithmic}
\end{algorithm}

\begin{algorithm}[tbh]
\caption{AugmentPredict}
\label{alg:augmentpredict}
\begin{algorithmic}[1]
\Require Classifier $f$, transform $\phi$ and parameter space $\Theta$, input object $x$, number of samples $n$, number of sample means $k$
\Ensure The set $Z\in\mathbb{R}^{k\times n}$ of realizations of random variable $Z$ from Eq.~\eqref{eq:prob_bound}
\For{$j \in [k]$}
\For{$i \in [n]$}
\State{$\theta \gets \theta \sim \Theta, \quad x' \gets \phi(x,\theta)$}
\State{$Z[j][i] \gets f(x')_2$}
\Comment{compute the value of $p'_2$ according to Eq. ~\eqref{eq:aug_pred}}
\EndFor
\EndFor
\State \Return $Z$
\end{algorithmic}
\end{algorithm}

The verification procedure against parametric transformations is described in Algorithms~\ref{alg:certification}-\ref{alg:augmentpredict}.

\subsection{Adaptation to generative models}

In the preceding sections, we considered robustness verification against parametric transformations that preserve the semantic class of the original audio signal. In this subsection, we extend the proposed framework to enable robustness verification of voice anti-spoofing models against generative models, specifically text-to-speech (TTS) and voice-cloning (VC) systems.

\subsubsection{TTS}
In the TTS setting, the objective is no longer to verify robustness with respect to transformations of a fixed input audio sample $x$, but rather to reason about an entire family (distribution) of audio signals generated by a TTS model. The generated audio depends primarily on the input text and, potentially, on additional generation parameters such as speech rate or selected artificial voice:
\begin{equation}
    \phi(x, \theta) \mapsto g(t, \theta),  
\end{equation}
where $g$ denotes the considered TTS model, $t \sim \mathcal{T}$, with $\mathcal{T} \subset \mathbb{R}^{l \times s}$ denoting a text representation of vocabulary size $s$ and length $l$, and $\theta \sim \Theta$ represents the collection of additional independent generation variables. Such variables may be multidimensional and either be fixed or drawn from the corresponding distributions; they can include, for example, speech rate, voice identity, or language configuration of the TTS system, in the case of multi-speaker or multilingual TTS. In practice, the text inputs are often drawn from natural language corpora or task-specific datasets.

Unlike sample-specific verification against transformations of the given input $x$, the key idea in this setting is to certify robustness with respect to the \emph{distribution of audio signals induced by the generative model $g$}. That is, we aim to quantify how frequently the classifier $f$ assigns a generated sample $x' = g(\theta)$ to the incorrect class of bona fide speech:
\begin{equation}
    \mathbb{P} [p_2' > 1/2].  
\end{equation}

\subsubsection{Voice cloning}
PV-VASM can be used to provide verification results for voice cloning (VC) systems. 
Voice cloning aims to synthesize speech for an arbitrary text $t$ while preserving the voice characteristics of a reference speaker. In a nutshell, a voice cloning model $g_{vc}$ can be expressed as
\begin{equation}
g_{\text{vc}}: \mathcal{T}  \times \mathcal{X} \times \Theta \to \mathcal{X}, \quad
x' = g_{\text{vc}}(t,  x_{\text{ref}}, \theta),
\end{equation}
where $t \in \mathcal{T}$ denotes the input text, $x_{\text{ref}} \in \mathcal{X}$ is a reference audio sample of the target speaker $s_{\text{tgt}}$, and $\theta \in \Theta$ represents additional generation parameters. This formulation naturally supports both sample-specific and distribution-level robustness verification. In the sample-specific setting, for a fixed input $x$ (which need not necessarily be bona fide speech), one can sample or segment the reference audio $x_{\text{ref}}$, as well as vary the input texts and additional generation parameters. In contrast, distribution-level verification further considers variability over the entire set of admissible input audio signals $x$, thereby assessing robustness with respect to the full data-generating distribution. Note that VC is a label-switching transformation for the bona fide inputs.

\begin{table*}[tb]
\caption{Robustness verification results against parametric input transformations. The desired confidence level is set to be $\alpha=10^{-6}$.}
\label{tab:main-table}
\resizebox{\textwidth}{!}{%
\centering
\small
\begin{tabular}{llllcrrrll}
\toprule
\multicolumn{1}{l}{\multirow{2}{*}{Transform}} &
\multirow{2}{*}{\begin{tabular}[c]{@{}l@{}}Parameters'\\ Space $\Theta$ \end{tabular}} &
\multicolumn{1}{l}{\multirow{2}{*}{$n$}} &
\multicolumn{1}{l}{\multirow{2}{*}{$k$}} &
\multicolumn{4}{c}{PCA} &
\multicolumn{1}{c}{\multirow{2}{*}{$\overline{\mathcal{A}(x)}$}} &
\multicolumn{1}{c}{\multirow{2}{*}{$\overline{p(x)}$}} \\
\cmidrule(lr){5-8}
&&&&
$\varepsilon=10^{-5}$ &
$\varepsilon=10^{-3}$ &
$\varepsilon=10^{-2}$ &
$\varepsilon=0.05$ &
& \\
\midrule
LPF & $\omega_{\text{max}}\in[2.5,3]~\mathrm{kHz}$ & 500 & 40 & 0.9767 & 0.9833 & 0.9900 & 0.9900 & 0.0049 & $8.3\cdot10^{-13}$ \\
LPF & $\omega_{\text{max}}\in[2.5,3]~\mathrm{kHz}$ & 1000 & 20 & 0.9767 & 0.9833 & 0.9900 & 0.9900 & 0.0048 & $2.0\cdot10^{-11}$ \\
\midrule
HPF & $\omega_{\text{min}}\in[0.5,1]~\mathrm{kHz}$ & 500 & 40 & 0.8433 & 0.8667 & 0.8733 & 0.8767 & 0.1070 & $7.5\cdot10^{-7}$ \\
HPF & $\omega_{\text{min}}\in[0.5,1]~\mathrm{kHz}$ & 1000 & 20 & 0.8433 & 0.8667 & 0.8733 & 0.8767 & 0.1065 & $3.6\cdot10^{-6}$ \\
\midrule
\begin{tabular}[c]{@{}l@{}}Background\\ Noise\end{tabular}
& $\mathrm{SNR}\in[15,30]~\mathrm{dB}$ & 500 & 12 & 0.3400 & 0.3467 & 0.3733 & 0.4300 & 0.1038 & 0.1687 \\
\begin{tabular}[c]{@{}l@{}}Background\\ Noise\end{tabular}
& $\mathrm{SNR}\in[15,30]~\mathrm{dB}$ & 1000 & 6 & 0.3467 & 0.3533 & 0.3867 & 0.4600 & 0.0943 & 0.1814 \\
\midrule
Pitch Shift & $\operatorname{ST}$ $ \in [-6,6]$ & 500 & 40 & 0.5867 & 0.6133 & 0.6133 & 0.6233 & 0.3506 & $1.4\cdot10^{-11}$ \\
Pitch Shift & $\operatorname{ST}$ $\in[-6,6]$ & 1000 & 20 & 0.5867 & 0.6133 & 0.6133 & 0.6233 & 0.3485 & $4.6\cdot10^{-11}$ \\
\midrule
BPF & $\omega \in [200,4000], b\in [0.5,1.99]$ & 500 & 40 & 0.0033 & 0.0033 & 0.0067 & 0.0100 & 0.3510 & 0.1664 \\
BPF & $\omega \in [200,4000], b \in[0.5,1.99]$ & 1000 & 20 & 0.0033 & 0.0033 & 0.0067 & 0.0200 & 0.3337 & 0.2114 \\
BPF & $\omega \in [200,1500], b\in  [1.2,1.5]$ & 500 & 40 & 0.9433 & 0.9533 & 0.9533 & 0.9533 & 0.0345 & $4.3\cdot10^{-13}$ \\
BPF & $\omega \in [200,1500], b \in[1.2,1.5]$ & 1000 & 20 & 0.9433 & 0.9533 & 0.9533 & 0.9533 & 0.0337 & $8.3\cdot10^{-13}$ \\
\midrule
Time Stretch & $\varkappa \in  [0.75,1.35]$ & 500 & 40 & 0.8067 & 0.8133 & 0.8300 & 0.8800 & 0.0404 & 0.0064 \\
Time Stretch & $\varkappa \in  [0.75,1.35]$ & 1000 & 20 & 0.8100 & 0.8233 & 0.8400 & 0.8900 & 0.0372 & 0.0077 \\
\midrule
Gain & $\gamma \in [-10,20]~\mathrm{dB}$ & 500 & 40 & 0.8800 & 0.8933 & 0.8933 & 0.8933 & 0.0384 & 0.0072 \\
Gain & $\gamma \in [-10,20]~\mathrm{dB}$ & 1000 & 20 & 0.8800 & 0.8933 & 0.8933 & 0.8933 & 0.0371 & 0.0060 \\
Gain & $\gamma \in [-10,10]~\mathrm{dB}$ & 500 & 40 & 0.9867 & 0.9867 & 0.9867 & 0.9867 & 0.0076 & $1.2\cdot10^{-15}$ \\
Gain & $\gamma \in [-10,10]~\mathrm{dB}$ & 1000 & 20 & 0.9867 & 0.9867 & 0.9867 & 0.9867 & 0.0075 & $1.5\cdot10^{-13}$ \\
\midrule
RIR & $r \in \text{OpenSLR dataset}$ & 1000 & 20 & 0.0800 & 0.0833 & 0.0866 & 0.1100 & 0.2523 & 0.2358 \\
\midrule
\begin{tabular}[c]{@{}l@{}}
Composite: \\
{[}Gain, LPF \\
Gaussian Noise{]}
\end{tabular} &
\begin{tabular}[c]{@{}l@{}}
 $\gamma \in [-10, 10]~\mathrm{dB}$ \\
 $\omega_{max} \in [2.5, 3]~\mathrm{kHz}$ \\
 $\sigma \in [0.01, 0.03]$
\end{tabular} & 1000 & 20 & 0.5733 & 0.5900 & 0.6167 & 0.6800 & 0.1140 & 0.0628\\
\bottomrule
\end{tabular}
}
\end{table*}

\section{Experimental setup}

\subsection{Source model, datasets, and hyperparameters}
We selected Wav2Vec2-AASIST as the architecture for the source model $f$. This model was trained for two epochs using cross-entropy loss and the AdamW~\cite{loshchilov2017decoupled} optimizer. 

For the train data, a combination of open-source datasets was used, including ASVspoof $19$, $21$ (LA and DF) splits, ASVspoof $5$,  ADD $22-23$, DFADD~\cite{du2024dfadd}, SONAR~\cite{li2024sonar}, CFAD~\cite{ma2024cfad}, MLAAD~\cite{muller2024mlaad}, Speech-to-Latex~\cite{korzh2025speech}, and Mozilla Common Voice~\cite{ardila2019common}.

During training, to improve the model's empirical robustness, each audio sample was subjected to a composition of randomly selected augmentations applied in a random order. Augmentations were chosen from the following set: voice activity detection, random signal cropping, background noise addition, room impulse response (RIR) simulation, Gaussian noise injection, band-, low-, and high-pass filter (BPF, LPF, HPF), loudness normalization, codec augmentation, time stretching, pitch shifting, bit crushing, and gain adjustment. For the test data, a balanced class subset of $300$ audio samples from the ASVspoof~5 evaluation (test) subset was used. Only initially correctly classified audio samples were used for the verification.

The following default values of hyperparameters were used in the experiments unless said otherwise: the value of  $\delta$ from Eq.~\eqref{eq:output} was set to $0.9$, the range of parameter $t$ from Eq.~\eqref{eq:prob_bound} was set to be $[-50, -10^{-4}]$. The number of transformations of input samples $n$ to compute the single statistic from Eq.~\eqref{eq:sample_mean}, the total number of statistics $k$, and an upper bound $\alpha$ for the error probability of the method from Eq.~\eqref{eq:gurantees} are varied for different experiments and are discussed in the subsequent sections.

\subsection{Parametric transformations and speech generation models}
We evaluate PV-VASM in two different settings, namely, against conventional input perturbations and against speech generation models.

In the first setting, we considered parametric perturbations \emph{that do not change the predicted class of the input object $x$}. For this purpose, several input transformations that are commonly applied as audio augmentation were used. In a nutshell, if the input audio is a deepfake, the method should verify that the classifier's prediction remains the same under an input transformation; analogously, if the input is a bona fide audio, the method should verify that the predicted class does not change under an input transformation. Audio transformations were implemented using  \texttt{audiomentations}~\cite{jordal2025iver56} and \texttt{torch-audiomentations}~\cite{jordal2022asteroid} libraries. If not stated otherwise, the default transformation parameters from the \texttt{audiomentations} library were used. 

Below we list the set of audio transformations used during the  training of the source model and indicate, where applicable, the range of parameters of a corresponding transformation used for augmentation of the source model during training:
\begin{itemize}
    \item \textbf{Filters}. The cutoff frequency $\omega_{max}$ for LPF is sampled from $[3400 \text{ Hz},7500\text{ Hz}]$; for the HPF, the cutoff frequency $\omega_{min}$ is sampled from $[100\text{ Hz}, 400\text{ Hz}]$; for the BPF, the cutoff frequency $\omega$ is sampled from $[200\text{ Hz}, 4000\text{ Hz}]$, and the bandwidth to central frequency ratio $b$ is sampled from $[0.5, 1.99]$.
    \item \textbf{Pitch shift}. Pitch was randomly adjusted from $-4$ to $4$ semitones (ST).
    \item \textbf{Time stretch}. A default speedup parameter $\varkappa$ was sampled from $[0.8, 1.25]$. 
    \item \textbf{Gain}. The gain parameter $\gamma$  was between $-10$ dB and $+20$ dB.
    \item \textbf{RIR}.     
    We use the OpenSLR dataset of RIR recordings~\cite{ko2017study} and sample uniformly from it. 
    \item \textbf{Background noise}. Similar to the RIR, we sampled from the noise subset of the Musan dataset~\cite{snyder2015musan}. 
\end{itemize}
In our experiments, the parameters of input transformations were sampled uniformly from the selected parameter's range or among available noise and RIR samples; however, all the parameters may be sampled from any other analytical distribution. We indicate that the ranges of input transformations during the verification procedure differ from the ones used during training, and are presented in the subsequent section. 

To evaluate the efficiency of the method against TTS and VC methods,  we generated data using the following open-source generative models:  Vosk\footnote{\url{https://pypi.org/project/vosk-tts}}, Silero\footnote{\url{https://pypi.org/project/silero-tts}},  Coqui XTTS-v2\footnote{\url{https://huggingface.co/coqui/XTTS-v2}}~\cite{casanova2024xtts}, f5-TTS~\cite{chen2025f5}, CosyVoice\footnote{\url{https://github.com/FunAudioLLM/CosyVoice}}, and proprietary models of ElevenLabs\footnote{\url{https://elevenlabs.io}} and Finevoice\footnote{\url{https://finevoice.ai}}. Texts used for the artificial annotation were primarily selected from the Mozilla Common Voice dataset.

\subsection{Metrics}
The primary quantitative measure of efficiency of a probabilistic verification method $\mathcal{A}$ in a classification problem is probabilistically certified accuracy \cite{pautov2022cc},  PCA, defined as 
\begin{align}
\operatorname{PCA}(\varepsilon, \alpha, \mathcal{D}) =  \frac{1}{|\mathcal{D}|}\sum_{(x,y) \in \mathcal{D}} \mathds{1}\{ & h(x)=y \wedge \mathcal{A}(x) <  \varepsilon \nonumber \\ & \wedge  p < \alpha /2 \},
\end{align}
where $p$ is from Algorithm \ref{alg:certification}. 
Informally, PCA is the fraction of objects from $\mathcal{D}$ which (i) are correctly classified by the source model $f$; (ii) are assigned the misclassification probability $\mathcal{A}(x)$ less than the threshold value $\varepsilon$; (iii) have the $p(x)$ is upper bounded by $\alpha/2$, given that $\tilde c$ is estimated with $1-\alpha/2$ confidence.

When verification is performed with respect to the TTS-induced distribution, the resulting PCA metric is binary valued and is defined as:
\begin{equation}
\label{eq:binary_pca}
    \operatorname{PCA}(\varepsilon, \alpha) = \mathds{1} \{\mathcal{A} \prec \varepsilon \wedge \ p \prec  \alpha/2 \},
\end{equation}
where $\mathcal{A}\prec \varepsilon$ means that $\mathcal{A}$ is uniformly less than $\varepsilon$.

\section{Results}
\subsection{Parametric input transformations}
In Table \ref{tab:main-table}, we present the results of robustness verification of the considered Wav2Vec2-AASIST model against parametric input transformations. For each transformation and corresponding space of parameters, we report the number of input transformations $n$ to compute the single statistic from Eq.~\eqref{eq:sample_mean} and the total number of statistics $k$; for the given value of threshold $\varepsilon,$ we present the values of PCA metric; additionally, we report the average value of $\overline{\mathcal{A}(x)}$ from Algorithm \ref{alg:certification} and the average value $\overline{p} = \overline{p}(n,k,\tilde{c}(x))$ from Eq. \eqref{eq:cc_cert_error}. 
\begin{figure}[tbh]
  \centering
  \includegraphics[width=\linewidth]{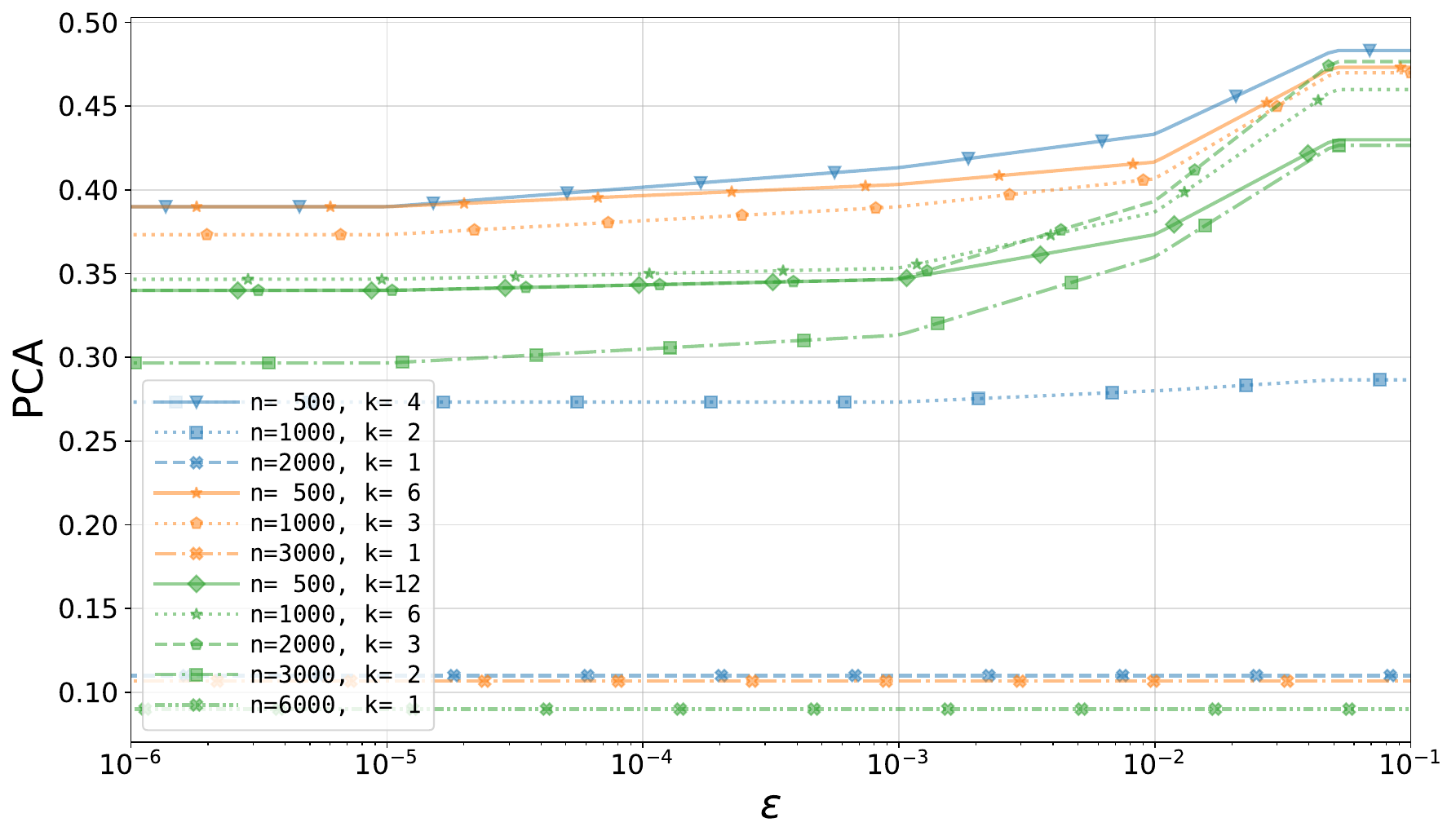}
  \caption{Dependence of PCA on $(m, n, k)$ for background noise perturbations with $\operatorname{SNR} \in [15,30]$. The confidence level is set to $\alpha=10^{-6}$. Curves sharing the same color correspond to an identical computational budget $m$, while line styles and marker types indicate variations in $n$ and $k$, respectively.}
  \label{fig:pca_vs_nk_bn}
\end{figure}

\begin{figure}[tbh]
  \centering
  \includegraphics[width=\linewidth]{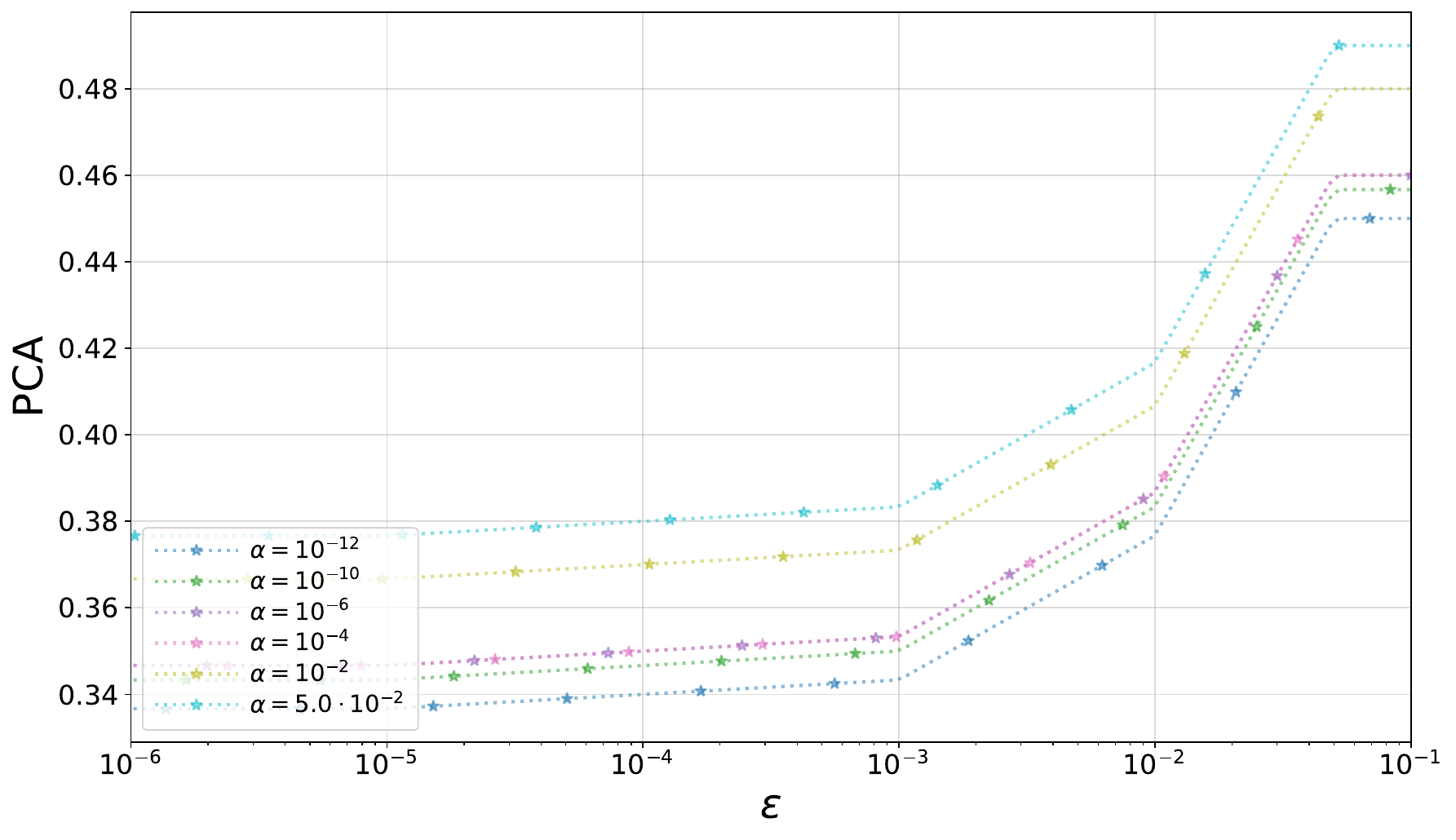}
  \caption{Dependence of PCA on $\alpha$ for background noise perturbations with $\operatorname{SNR} \in [15,30]$. $m=6000,~n=1000,~k=6$ are fixed.}
  \label{fig:pca_vs_alpha_bg}
\end{figure}
\begin{figure}[tbh]
  \centering
  \includegraphics[width=\linewidth]{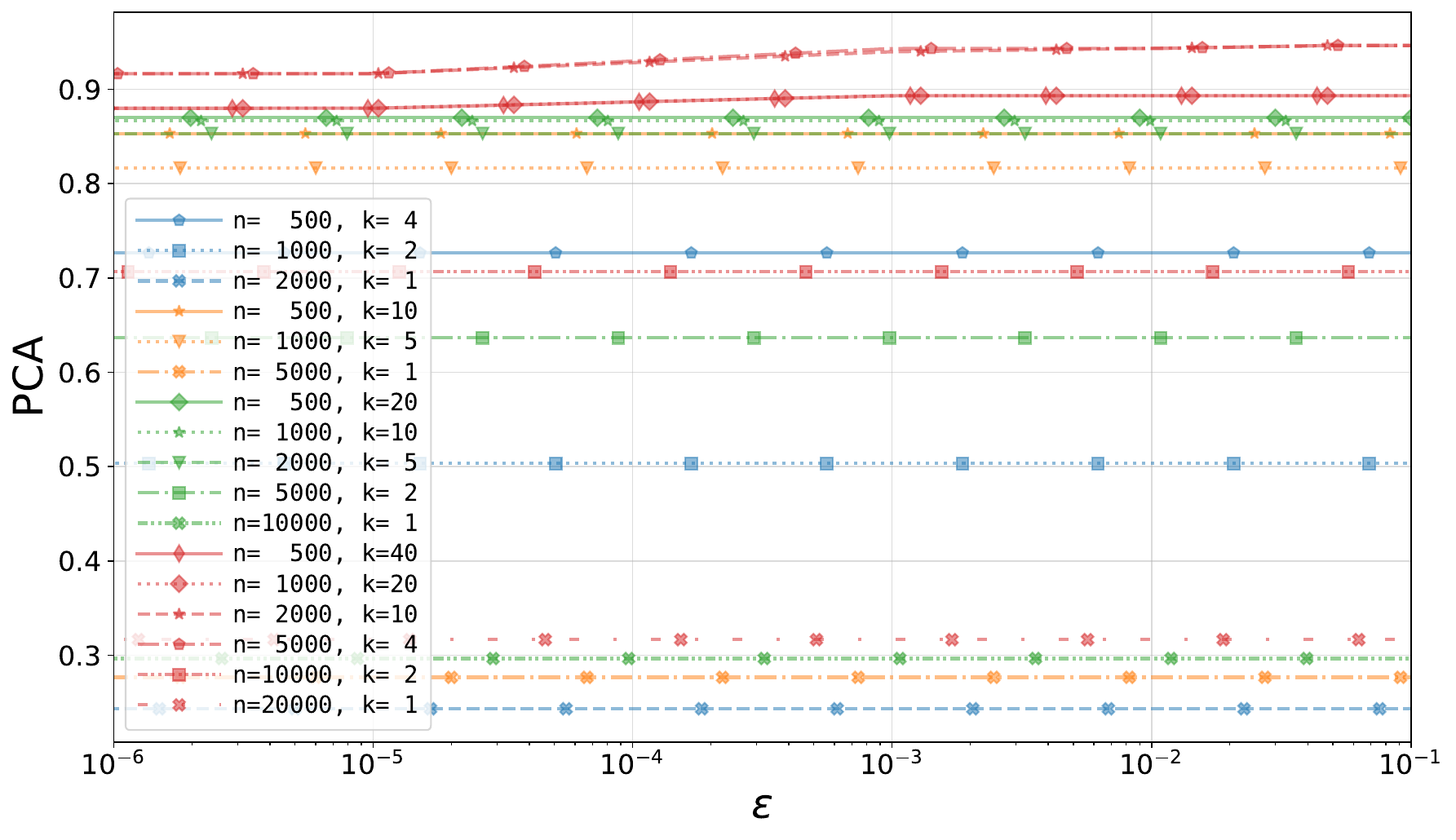}
  \caption{Dependence of PCA on $(m, n, k)$ for the gain adjustment transform with $\gamma \in [-10,20]~\operatorname{dB}$. The confidence level is set to $\alpha=10^{-6}$. Curves sharing the same color correspond to the same augmentation budget $m$, while line styles and marker types indicate variations in $n$ and $k$, respectively.}
  \label{fig:pca_vs_nk_gain}
\end{figure}
\begin{figure}[tbh]
  \centering
  \includegraphics[width=\linewidth]{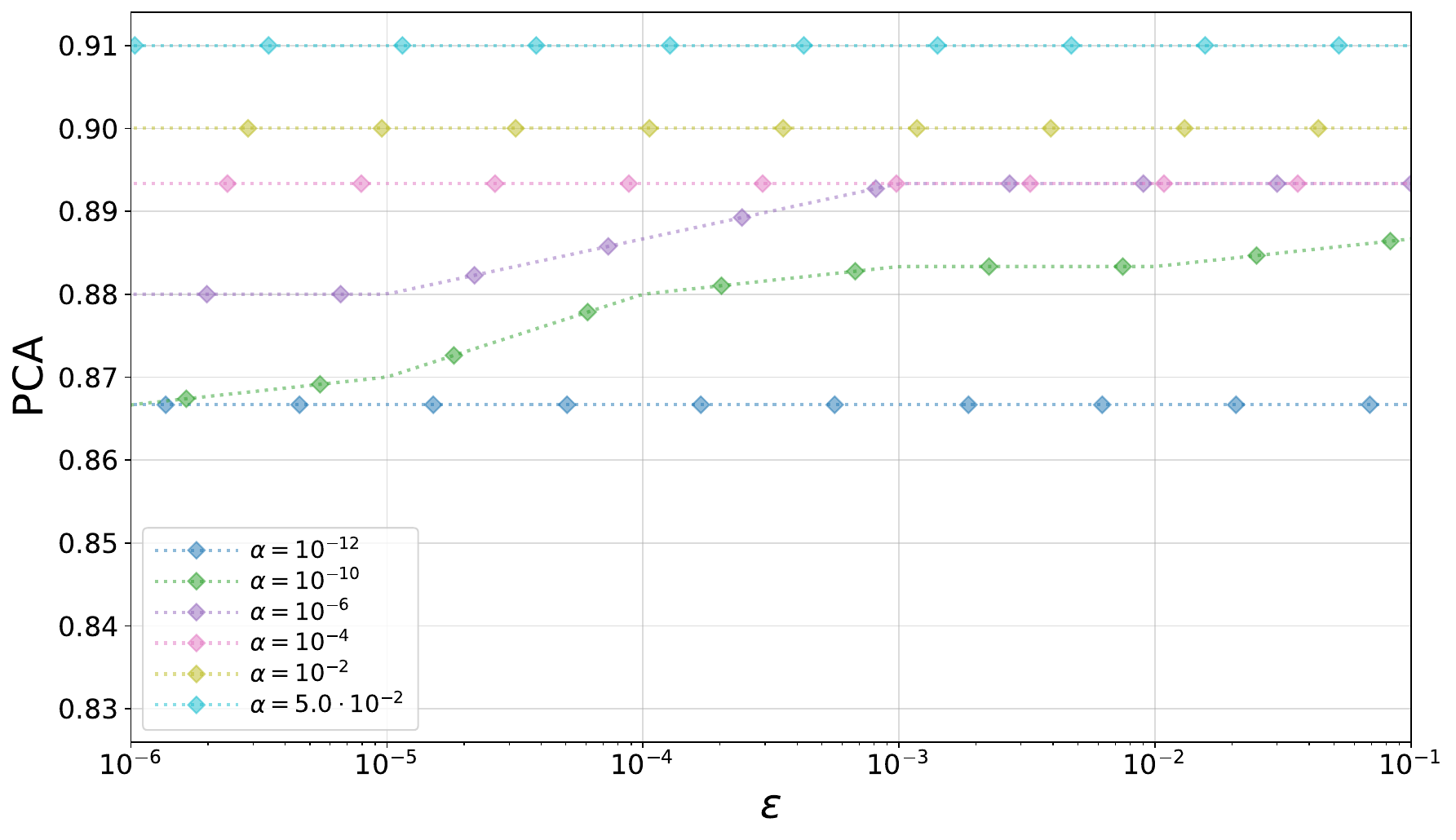}
  \caption{Dependence of PCA on $\alpha$ for the gain adjustment transform with  $\gamma \in [-10,20] \operatorname{dB}$. The values $m=20000,~n=1000,~k=20$ are fixed.}
  \label{fig:pca_vs_alpha_gain}
\end{figure}
From the results, we make several observations. Firstly, the source model show strong robustness against LPF, HPF, and time stretch transforms, demonstrating low average error probability $\overline{p}$; secondly, the different distributions of the total augmentation budget in the form $m=n\times k$ into $k$ splits of $n$ samples each (see, for example, Eq. \eqref{eq:sample_mean}) do not always affect the verification results, but can be used to control the average error probability; finally, we highlight that the broader the parameters' space $\Theta$ for the given transform, the worse the expected robustness of the model to that transform is (see, for instance, the verification results for BPF for different $\Theta$). 

\begin{figure}[tbh]
  \centering
  \includegraphics[width=\linewidth]{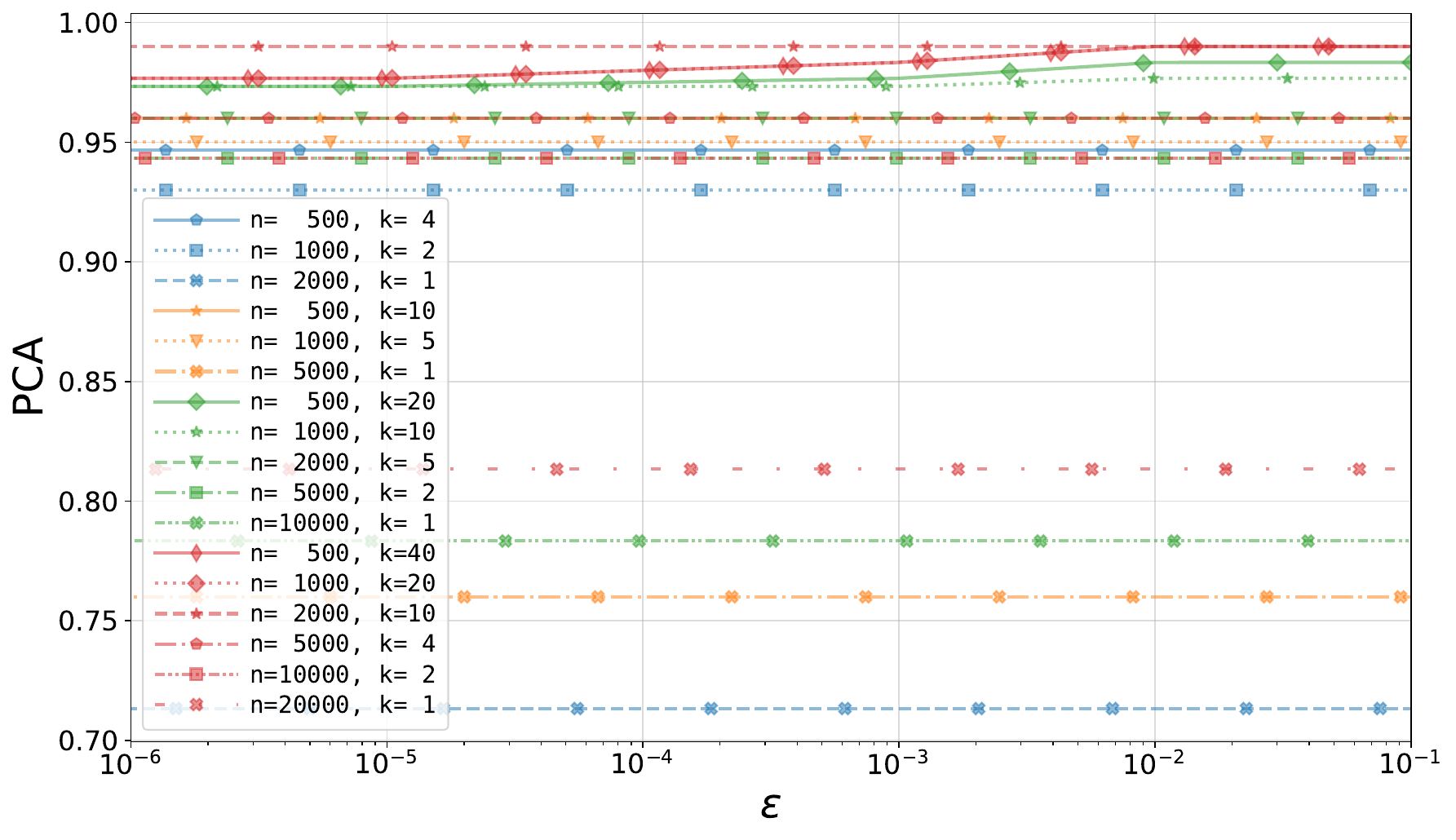}
  \caption{Dependence of PCA on $(m, n, k)$ for the low pass filter with the cutoff frequency  $\omega_{max}$ is randomly sampled from $[2500, 3000] \operatorname{Hz}$ range. The confidence level is set to $\alpha=10^{-6}$.  Curves sharing the same color correspond to the same augmentation budget $m$, while line styles and marker types indicate variations in $n$ and $k$, respectively.}
  \label{fig:pca_vs_nk_lpf}
\end{figure}

\begin{figure}[tbh]
  \centering
  \includegraphics[width=\linewidth]{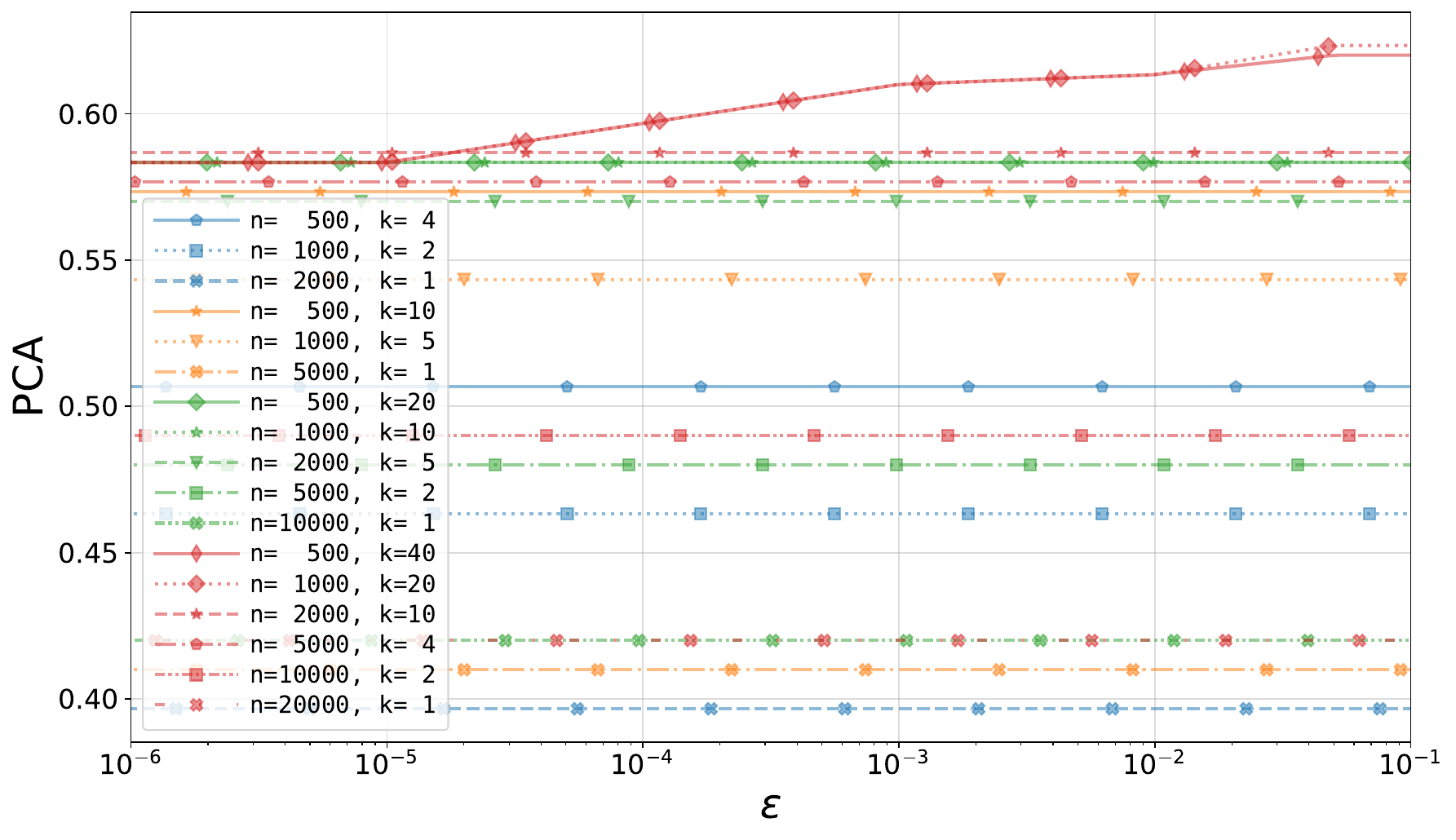}
  \caption{Dependence of PCA on $(m, n, k)$ for the pitch shift transform,  ST $\in[-6, 6]$ semitones. The confidence level is set to be $\alpha=10^{-6}$. Curves sharing the same color correspond to the same augmentation budget $m$, while line styles and marker types indicate variations in $n$ and $k$, respectively.}
  \label{fig:pca_vs_nk_ps}
\end{figure}
In Figures \ref{fig:pca_vs_nk_bn}, \ref{fig:pca_vs_nk_gain}, \ref{fig:pca_vs_nk_lpf}, \ref{fig:pca_vs_nk_ps}, we report the dependency of PCA on the values of augmentation budget $m$ and its distribution; in Table ~\ref{tab:pca_on_nk_lpf}, we report averaged probability bounds from Algorithm \ref{alg:certification} along with the averaged error probability  $\overline{p}(x)$ for specific transforms. In Figures~\ref{fig:pca_vs_alpha_bg} and \ref{fig:pca_vs_alpha_gain}, we compare PCA for the different values of  $\alpha$ given fixed $(m,n,k)$.
\begin{table}[tbh]
\caption{Dependency of average probability of error $\overline{p}$ for different values $(n, k, m),$ low pass filter, $\Theta  = [2500,~3000]\operatorname{Hz}$. The confidence level is set to $\alpha=10^{-6}$.}
\label{tab:pca_on_nk_lpf}
\centering
\begin{tabular}{l l l l l }
\toprule
$n$ & $k$ & $m$ &  $\overline{\mathcal{A}(x)}$          & $\overline{p(x)}$ \\ 
\midrule
500 & 4 & 2000 &  $4.7\cdot 10^{-3}$      & $1.3\cdot 10^{-3}$ \\ 
1000 & 2 & 2000 &  $4.8\cdot 10^{-3}$      & $2.3\cdot 10^{-3}$ \\ 
2000 & 1 & 2000 &  $4.6\cdot 10^{-3}$      & $5.0\cdot 10^{-3}$ \\ 
\midrule
500 & 10 & 5000 & $4.8\cdot 10^{-3}$        & $2.9\cdot 10^{-5}$ \\ 
1000 & 5 & 5000 &  $4.7\cdot 10^{-3}$       & $7.2\cdot 10^{-5}$ \\ 
5000 & 1 & 5000 &   $4.7\cdot 10^{-3}$      & $2.1\cdot 10^{-3}$ \\ 
\midrule
500 & 20 & 10000 &  $4.9\cdot 10^{-3}$       & $7.6\cdot 10^{-8}$ \\ 
1000 & 10 & 10000 & $4.8\cdot 10^{-3}$       & $3.8\cdot 10^{-7}$ \\ 
2000 & 5 & 10000 &  $4.7\cdot 10^{-3}$      & $4.3\cdot 10^{-6}$ \\ 
5000 & 2 & 10000 &   $4.7\cdot 10^{-3}$      & $1.2\cdot 10^{-4}$ \\ 
10000 & 1 & 10000 &  $4.7\cdot 10^{-3}$       & $1.0\cdot 10^{-3}$ \\ 
\midrule
500 & 40 & 20000 &  $4.9\cdot 10^{-3}$       & $8.3\cdot 10^{-13}$ \\ 
1000 & 20 & 20000 &  $4.8\cdot 10^{-3}$       & $2.0\cdot 10^{-11}$\\ 
2000 & 10 & 20000 &  $4.8\cdot 10^{-3}$       & $1.8\cdot 10^{-9}$ \\ 
5000 & 4 & 20000 &  $4.7\cdot 10^{-3}$       & $7.2\cdot 10^{-7}$ \\ 
10000 & 2 & 20000 & $4.7\cdot 10^{-3}$        & $3.1\cdot 10^{-5}$\\
20000 & 1 & 20000 &  $4.7\cdot 10^{-3}$       & $5.2\cdot 10^{-4}$\\ 
\bottomrule
\end{tabular}
\end{table}

\subsection{TTS}

\begin{table}[tbh]
\caption{Robustness verification results against TTS generators. Results are shown for the model before and after finetuning on these generators. The confidence level is set to be $\alpha=10^{-6}$.}
\label{tab:tts}
\resizebox{\columnwidth}{!}{%
\centering
\small
\begin{tabular}{llllll}
\toprule
TTS model & $n$ & $k$ & $\delta$ & ${\mathcal{A}}$ & ${p}$ \\
\midrule
\multicolumn{6}{c}{Pre-finetuning results} \\ 
\midrule
Vosk & 5000 & 26 & 0.75 & 0.1352 & $3.37 \cdot 10^{-17}$ \\
Silero & 650 & 20 & 0.5 & 0.5030 & $6.1\cdot10^{-21}$ \\
CosyVoice & 500 & 40 & 0.9 & 0.5236 & $2.4\cdot10^{-8}$ \\
f5 & 10000 & 20 & 0.9 & 0.4019 & $2.6\cdot10^{-25}$ \\ 
ElevenLabs & 500 & 6 & 0.5 & 0.3308 & $3.0\cdot10^{-5}$ \\ 
Finevoice & 200 & 25 & 0.5 & 0.1446 & $3.8\cdot10^{-7}$ \\ 
\midrule
\multicolumn{6}{c}{Post-finetuning results} \\
\midrule
Vosk & 5000 & 26 & 0.75 & 0.0686 & $8.1\cdot10^{-7}$ \\
Silero & 650 & 20 & 0.5 &  0.0579 & $1.64\cdot10^{-5}$ \\
CosyVoice & 500 & 40 & 0.9 & 0.2656 & $3.2\cdot10^{-4}$ \\ 
f5 & 10000 & 20 & 0.9 & 0.2319 & $1.7\cdot10^{-16}$ \\
ElevenLabs & 500 & 6 & 0.5 & 0.2002 & $3.9\cdot10^{-3}$ \\ 
Finevoice & 200 & 25 & 0.5 & 0.0721 & $1.7\cdot10^{-5}$ \\ 
\bottomrule
\end{tabular}
}
\end{table}

\begin{figure}[tbh]
  \centering
  \includegraphics[width=\linewidth]{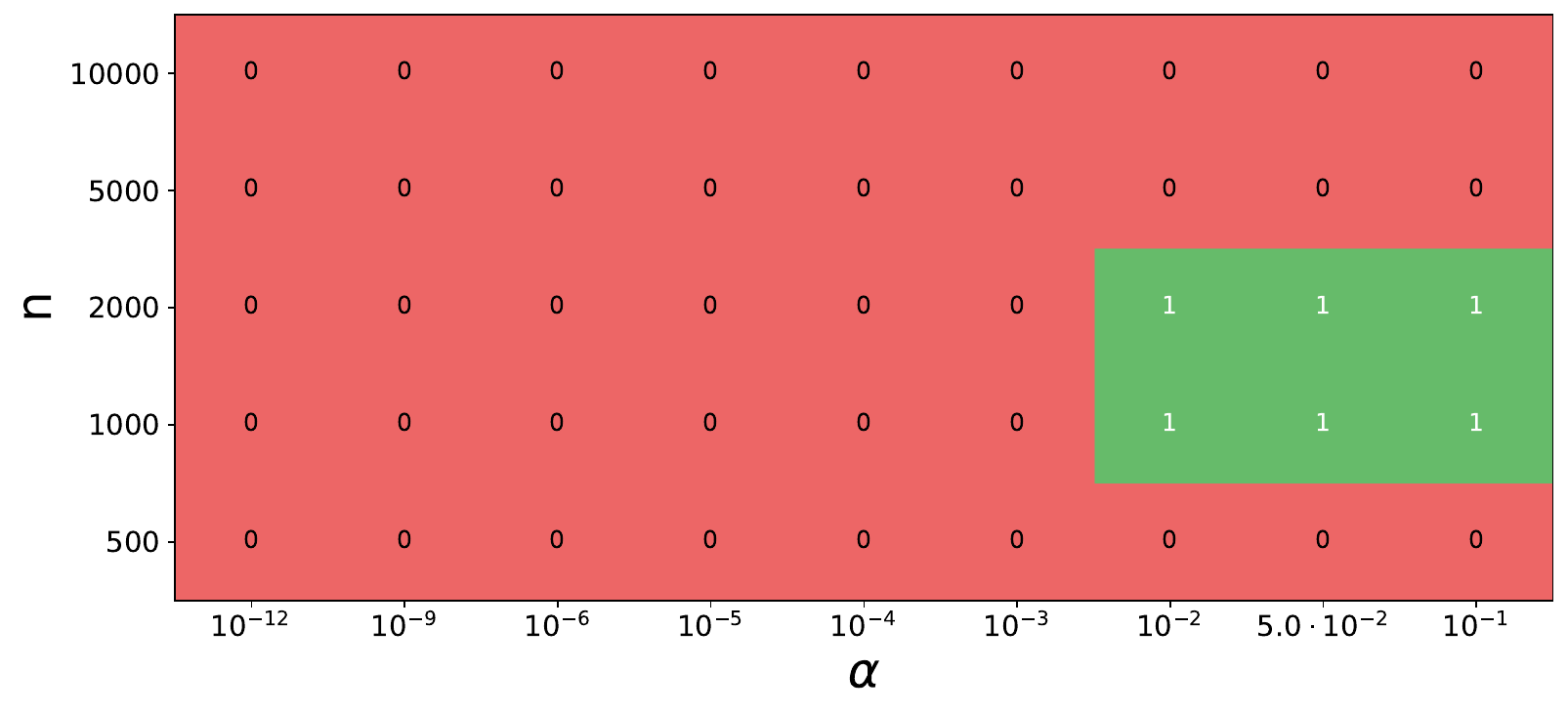}
  \caption{Verification condition result from Eq.~\eqref{eq:binary_pca} for the pre-finetuned $f$ against Vosk TTS vs. various $\alpha$ and $n$ given fixed $m=130000$, $\delta=0.9$, and $\varepsilon=0.1$.}
  \label{fig:vosk_tts}
\end{figure}

In Table~\ref{tab:tts}, the results of robustness verification against different TTS generators are shown.  Here, in the column ``$p$'', an estimation of the method's error probability from Eq.~\eqref{eq:cc_cert_error} is presented. We highlight that in this setting, the verification of the VAS model's robustness against the \emph{distribution} of perturbations induced by a TTS model is performed, so the output of the algorithm and its error probability depend on the distribution. We present verification results for the pre- and post-finetuned models. Here,  the pre-finetuned model is the base model trained on the original dataset; the post-finetuned model represents the base model that was additionally trained on a subset of the data generated by a corresponding TTS model. For each TTS model, all the generated samples were split into a train subset ($10\%$) and the verification subset ($90\%$). In Figure~\ref{fig:vosk_tts}, we present the results of the verification of robustness against the Vosk TTS model. To do so, 
we fixed the augmentation budget $m=130000$ and the probability threshold $\varepsilon=0.1$ and show for which values of $(n,\alpha)$  the verification condition from Eq.\eqref{eq:binary_pca} is met. It is noteworthy that the verification condition is met for a few pairs of $(n,\alpha)$ due to the trade-off between  $\mathcal{A}$ and $p$  for different values of $n:$ on the one hand, the larger the $n$, the smaller the value of $\mathcal{A}$; on the other hand, the larger the $n$, the larger the value of $p$. From the results, the following observations are made. First, the verification of robustness against TTS is a more complicated task than the verification of robustness against random input transformations (one can observe, for example, the range of $\mathcal{A}$ in the TTS experiment, making the verification impossible for small values of $\varepsilon$). Secondly, fine-tuning on data obtained from the same TTS expectedly improves the verification results (up to $1.5 -3$ times in terms of values of $\mathcal{A}$).

\subsection{Voice Cloning}
Similarly to the TTS experiment, we measured the model's robustness against VC models. We present the results of verification of robustness against  XTTSv2 and F5 models in Table~\ref{tab:vc}. 
One can notice that the improvement of PCA for the XTTSv2 model after finetuning is slightly less than the one for the f5 model; this can be explained by the fact that some inclusion of audio samples generated by the XTTSv2 model is present in the training dataset of the base model $f$.

\begin{table}[tbh]
\caption{Robustness verification results against VC generators. Results are shown for the model before and after finetuning on these generators. Confidence level is set to $\alpha=10^{-4}$, $\delta=0.5$.}
\label{tab:vc}
\resizebox{\columnwidth}{!}{%
\centering
\begin{tabular}{lllcrrrll}
\toprule
\multicolumn{1}{l}{\multirow{2}{*}{VC model}} &
\multicolumn{1}{l}{\multirow{2}{*}{$n$}} &
\multicolumn{1}{l}{\multirow{2}{*}{$k$}} &
\multicolumn{3}{c}{PCA} &
\multicolumn{1}{c}{\multirow{2}{*}{$\overline{\mathcal{A}(x)}$}} &
\multicolumn{1}{c}{\multirow{2}{*}{$\overline{p(x)}$}} \\
\cmidrule(lr){4-6}
&&&
$\varepsilon=0.05$ &
$\varepsilon=0.10$ &
$\varepsilon=0.30$ &
& \\
\midrule
\multicolumn{8}{c}{Pre-finetuning results} \\
\midrule
XTTSv2 & 100 & 50 & 0.92 & 0.94 & 1.00 & 0.0148 & $3.8\cdot10^{-6}$ \\
f5 & 500 & 10 & 0 & 0 & 0.24 & 0.4167 & $9.6\cdot10^{-5}$ \\
\midrule
\multicolumn{8}{c}{Post-finetuning results} \\
\midrule
XTTSv2 & 100 & 50 & 0.96 & 0.96 & 1.00 & 0.0078 & $4.1\cdot10^{-6}$ \\
f5 & 500 & 10 & 0.00 & 0.00 & 0.30 & 0.3531 & $6.8\cdot10^{-4}$ \\
\bottomrule
\end{tabular}
}
\end{table}

\subsection{Optimal values of hyperparameters}
\label{subsec:var}

Recall that the PV-VASM yields an upper bound for the probability of misclassification of the audio subject to the transformation from  Eq.~\eqref{eq:second_main_prob}. This upper bound may be over-conservative, i.e., significantly higher than the actual unknown value of $\mathbb{P}_{\theta \sim \Theta}[h(x) \ne h(x')]$. We also highlight that the distribution of the augmentation budget (namely, $m=n\times k$) into $k$ chunks of $n$ samples each affects the upper bound for error probability (see, for example, Eq.~\eqref{eq:cc_cert_error}). To study these dependencies, we conduct additional experiments to illustrate the effect of the computational budget re-distribution. 

Namely, given a fixed budget $m$, we compute the value of PCA and error probability for different pairs $(n_i, k_i)\colon n_i \times k_i = m.$ We present the PCA results for LPF in Fig.~\ref{fig:pca_vs_nk_lpf} and Table~\ref{tab:pca_on_nk_lpf}, for the background noise in Fig.~\ref{fig:pca_vs_nk_bn}, gain in Fig.~\ref{fig:pca_vs_nk_gain}, and pitch shift in Fig.~\ref{fig:pca_vs_nk_ps}. It can be observed that not only does a budget impact PCA. For instance, $k = 1$ uniformly yields the worst PCA results, and generally increasing $k$ improves verification results for a fixed $m$,  as it improves values of $p(x)$. However, it is not always satisfied, for example, see Fig.~\ref{fig:pca_vs_nk_bn}. To obtain a tighter bound, one would be recommended to balance the augmentation budget towards increasing the value of $k.$

In Figures~\ref{fig:pca_vs_alpha_bg} and~\ref{fig:pca_vs_alpha_gain},  the dependency of verification results on the confidence level $\alpha$ for background noise and gain transforms is illustrated. Generally, a higher value of $\alpha$ leads to less strict verification conditions and larger PCA. Additionally, for the TTS, we studied the dependency of  PCA  on values of $n$ and $\alpha$, with the other parameters fixed. The overall verification results are shown in Fig.~\ref{fig:vosk_tts}. It is worth mentioning that the lower the value of $n$, the lower the bound on the error probability ${p},$ but the looser the algorithm output $\mathcal{A}.$ To obtain a tighter verification results, a balance between the values of $p$ and $\mathcal{A}$ should be found.

\section{Discussion and limitations} 

For relatively simple input transformations, such as LPF with high cutoff frequency $\omega_{\text{max}}$ or moderate gain, good verification results are obtained. In contrast, robustness degrades for harder transformations that significantly reduce speech intelligibility, such as strong background noise, narrow high-band BPF, or combinations of several perturbations.

For robustness verification against TTS and VC generators, we again observe limited generalization of VAS models, reflected by poor upper bounds on the misclassification probability. The results also clearly show improved robustness after finetuning on these domains. Intuitively, when the distribution of the random variable $e^{tZ}$ from Eq.~\eqref{eq:prob_bound}, induced for example by a speech generation model, has a high variance within the support, the base model does not show a satisfactory level of robustness. In our experiments, this manifests as overly conservative PCA values for complex conventional and generative transformations. Although increasing the augmentation budget reduces the gap between the true misclassification probability $\mathbb{P}_{\theta \sim \Theta}[h(x) \ne h(x')]$ and its estimate $\mathcal{A}(x)$, it can be difficult to distinguish between overly conservative estimates and genuinely poor robustness.

The tightness of the bound in Eq.~\eqref{eq:output} and the error probability in Eq.~\eqref{eq:error} both depend on the parameter $t$ and hyperparameter $\delta$, and this trade-off should be considered when selecting (sub)optimal values. In our implementation, $t$ is restricted to a bounded range from Eq.~\eqref{eq:prob_bound}, while the optimal value may lie outside this interval; in general, the wider the range, the better verification results are expected.

Finally, the upper bound for error probability from Eq.~\eqref{eq:gurantees}, namely, $\alpha$, is equally distributed over (i) interval estimation of the coefficient of variation from Eq.~\eqref{eq:c_error} and (ii) estimation of error the verification algorithm from Eq.~\eqref{eq:cc_cert_error}; an increase of the confidence level in  Eq.~\eqref{eq:c_error} may positively affect the verification results. Additionally, while the classification threshold is set to $1/2$ according to Eq.~\eqref{eq:clf}, one could possibly adapt the verification methodology to a variable classification threshold to balance between false negative and false positive error rates.

\section{Conclusion and future work}

In this paper, we proposed PV-VASM, a framework for robustness verification of voice anti-spoofing models. We theoretically derived an upper bound for the error probability of the method and experimentally demonstrated the verification results in different settings, including the presence of parametric input perturbations, text-to-speech generation models, and voice cloning methods.  
We showed that the robustness of voice anti-spoofing models crucially depends on the type of perturbation and the width of its parameter space. We therefore confirmed that robustness to a simple parametric perturbation is noticeably higher than that to perturbations produced by a speech generation model. The proposed method can be applied to verify the robustness of models before real-world applications. 
Future work might be focused on tightening error bounds and adapting the proposed approach to spoofing-aware speaker verification methods.

\section{Generative AI use disclosure}
AI models (ChatGPT) and tools (Grammarly) were used only for text polishing and shortening.

\bibliographystyle{IEEEtran}
\bibliography{mybib}

\end{document}